\input harvmac.tex

\def\IL{\relax{\rm I\kern-.18em L}}
\def\IP{\relax{\rm P\kern-.18em P}}
\def\inbar{\,\vrule height1.5ex width.4pt depth0pt}
\def\IB{\relax{\rm I\kern-.18em B}}
\def\IC{\relax\hbox{$\inbar\kern-.3em{\rm C}$}}
\def\ID{\relax{\rm I\kern-.18em D}}
\def\IE{\relax{\rm I\kern-.18em E}}
\def\IF{\relax{\rm I\kern-.18em F}}
\def\IG{\relax\hbox{$\inbar\kern-.3em{\rm G}$}}
\def\IH{\relax{\rm I\kern-.18em H}}
\def\II{\relax{\rm I\kern-.18em I}}
\def\IK{\relax{\rm I\kern-.18em K}}
\def\IL{\relax{\rm I\kern-.18em L}}
\def\IM{\relax{\rm I\kern-.18em M}}
\def\IN{\relax{\rm I\kern-.18em N}}
\def\IO{\relax\hbox{$\inbar\kern-.3em{\rm O}$}}
\def\IP{\relax{\rm I\kern-.18em P}}
\def\IQ{\relax\hbox{$\inbar\kern-.3em{\rm Q}$}}
\def\IR{\relax{\rm I\kern-.18em R}}
\font\cmss=cmss10 \font\cmsss=cmss10 at 7pt
\def\IZ{\relax\ifmmode\mathchoice
{\hbox{\cmss Z\kern-.4em Z}}{\hbox{\cmss Z\kern-.4em Z}}
{\lower.9pt\hbox{\cmsss Z\kern-.4em Z}}
{\lower1.2pt\hbox{\cmsss Z\kern-.4em Z}}\else{\cmss Z\kern-.4em Z}\fi}
\def\IGa{\relax\hbox{${\rm I}\kern-.18em\Gamma$}}
\def\IPi{\relax\hbox{${\rm I}\kern-.18em\Pi$}}
\def\ITh{\relax\hbox{$\inbar\kern-.3em\Theta$}}
\def\IOm{\relax\hbox{$\inbar\kern-3.00pt\Omega$}}

\def\CP {{\cal P }}

\def\CV {{\cal V}}
\def\CO {{\cal O}}

\def\p {\partial}

\def\sdtimes{\mathbin{\hbox{\hskip2pt\vrule height 4.1pt depth -.3pt width
.25pt
\hskip-2pt$\times$}}}

\def\c{\cdot}

\Title{ \vbox{\baselineskip12pt\hbox{hep-th/9404025}\hbox{YCTP-P1-94}
\hbox{RU-94-19}}}
{\vbox{
\centerline{Addendum to: }
\centerline{Symmetries of the Bosonic String}
\centerline{S-Matrix} }}
\bigskip
\centerline{Gregory Moore\foot{ On leave of absence from Yale University. }
}
\bigskip
\centerline{moore@castalia.physics.yale.edu}
\smallskip\centerline{Dept. of Physics, Yale University,New Haven, CT}
\centerline{and}
\centerline{moore@charm.rutgers.edu}
\centerline{Dept. of  Physics, Rutgers University, Piscataway, NJ}
\bigskip
\bigskip

\noblackbox

\noindent
In a previous paper
we showed that bracket
relations uniquely fix the tree-level bosonic string
$S$-matrix for $N\leq 26$ particle scattering.
In this note we extend the proof to $N$-particle
scattering for all $N$.
\bigskip

\Date{January 19, 1994}


\newsec{Introduction}

In a previous paper
\ref\bracket{
G. Moore, ``Symmetries of the bosonic string $S$-matrix,''
hep-th/9310026}
we argued that a certain
set of relations on bosonic string $S$-matrix elements,
which we called bracket relations, are an expression
of spontaneous symmetry breaking of large underlying
symmetry algebras in string theory. Moreover,  we
argued that these relations
uniquely determine the tree-level bosonic string $S$-matrix
up to a choice of string coupling.

The latter claim is a mathematically precise statement
capable of rigorous proof. A technical point  restricted our
discussion to $N$-particle scattering for $N\leq 26$.
In this note we overcome this technical point with a
simple trick and extend the discussion
of \bracket\
to $N$ -particle scattering for all $N$.

\newsec{Uniqueness theorem}

Let us recall the main statement of \bracket.
Let $\CC_1$ be the CFT for the open bosonic string
in $\IR^{1,25}$. We will allow momenta and polarization
tensors to be complex.  Let $\CH$ be the ghost number 1
BRST cohomology.  $\CH$  is a representation of  the
complexified Poincar\'e group
$\CP_{26}\equiv  \IC^{26}\sdtimes O(26,\IC)$.
If  a linear function $\CA: \CH^{\otimes N} \to \IC$
satisfies three axioms and the ``bracket relations''
then it is uniquely
determined up to an overall constant $c_N$.
The axioms are:
P1: {\it Poincar\'e invariance}.
P2: {\it Analytic structure}.  As an analytic function
 $\CA$ is  meromorphic in the invariants  $s_{ij}=p_i\cdot p_j$.
It  has poles in $s_{ij}$ $\Leftrightarrow$
$p_I^2\in \{ 2,0,-2,-4,\dots \}$ for some momentum $p_I$ in an
intermediate channel in some dual diagram. $\CA$ is a
polynomial in the remaining relativistic invariants.
P3: {\it Regge Growth. } In the limit that  some
 $s_{ij}\to \infty$, holding other independent
invariants  fixed, the amplitude behaves as
$\CA \sim \alpha s_{ij}^\beta$.

The bracket  relations  state that
if  $J$ is a ghost number 1 BRST class of momentum $q$
and if $V_i$ are BRST classes of momenta $p_i$
such that $J$ is mutually local w.r.t all the $p_i$
(true iff $q\cdot p_i\in\IZ$ in the standard background), and if
$q+\sum p_i=0$  we have an identity
\eqn\genfdr{
\sum_i (-1)^{q\c p_2+\cdots q\c p_i}
\CA\bigl(V_1,\dots, \{ J, V_i \} ,\dots V_N\bigr) =0
}
where we have introduced the ``on-shell BV bracket''
\ref\lziii{B. Lian and G. Zuckerman,
``New perspectives on the brst algebraic structure of string theory,''
(hep-th/9211072) Commun. Math. Phys. {\bf 154}613 (1993).}
\eqn\osbv{
\{ \CO_1, \CO_2\} = \oint_z dw \bigl(b_{-1} \CO_1(w) \bigr) \CO_2(z)
\qquad .
}

In \bracket\ we were only able to prove our theorem
for $N\leq 26$ for the following reason.

\newsec{Problem}

First recall a few facts about $M$-particle scattering in $d$ dimensions.
Suppose $p_i\in \IC^d, i=1,\dots M$ satisfy $\sum_{i=1}^M p_i=0$.
Consider the $O(d)$ invariants $s_{ij}= p_i\cdot p_j$. These are not all
algebraically
independent, both because the momenta add to zero, and, possibly,
because the $p_i$ are confined to  $d$-dimensions.
The number of algebraically independent  invariants $s_{ij}$
is:
\eqn\nbrinv{
\eqalign{
Md -\half d(d+1) & \qquad\qquad {\rm for}\quad  M\geq d\cr
\half M(M-1) & \qquad\qquad {\rm for}\quad  M\leq d+1\cr}
}
For $M\leq d$ a set of algebraically independent
invariants can be chosen to be, for example,
$s_{ij}$ for $1\leq i\leq j\leq M-1$.
For $M\geq d+1$
the story is more complicated. For example,
since any set of $d+1$ momenta is linearly
dependent,
$\det \Delta=0$ where $\Delta$ is any $(d+1)$-dimensional
minor of the matrix $(s_{ij})$. Geometrically, the $s_{ij}$ must
lie on a variety $\CV[M,d]\subset \IC^{M(M+1)/2}$
of dimension \nbrinv.
We will refer to  invariants $s_{ij}$ where $i\not=j$ as
``kinematic invariants.''  (For example, we might take these to be
 $s,t$ for $4$-particle scattering.).

Writing bracket relations requires finding special configurations
of momenta.  Given such configurations one uses Lorentz invariance
and analyticity to deduce identities for all scattering configurations.
In particular, to write a bracket relation for $N$-particle scattering
amplitudes in $d$ noncompact dimensions we must
find momenta $p_1,\dots, p_N\in \IC^d$
such that, if we define $q\equiv -\sum p_i$ then
\eqna\conds
$$\eqalignno{p_i^2  &=2-2n_i &\conds a\cr
(p_i+q)^2 =\bigl(\sum_{j:j\not= i} p_j\bigr)^2&=2-2\tilde n_i
&\conds b\cr
p_i\c p_j &= z_{ij}&\conds c\cr}$$
where the levels
$n_i$, $\tilde n_i$, $i=1,\dots N$, and an algebraically
independent set of kinematic invariants for $N$-particle
scattering,  $z_{ij}$ ($i\not=j$), have been specified. The number
of available independent variables in \conds\ is the
number of invariants for $N+1$ particle scattering. Therefore
we may compare the number of equations and the number
of independent variables as follows:

\bigskip

\def\tablerule{\omit&\multispan{6}{\tabskip=0pt\hrulefill}&\cr}
\def\tablepad{\omit&height3pt&&&&&&&\cr}
$$\vbox{\offinterlineskip\tabskip=0pt\halign{
\strut$#$\quad&\vrule#&\quad\hfil $#$ \hfil\quad &\vrule #&\quad \hfil $#$
\hfil \quad&\vrule #& \quad $#$ \hfil\ &\vrule#&\quad $#$\cr
&\omit&\hbox{ }&\omit&\hbox{ $N\leq d$}&\omit&
\hbox{ $N>d$ }&\omit&\cr
\tablerule\tablepad
&& \#  {\rm Equations}&& \half N(N+1) 
&& Nd + N - \half d(d+1) && \cr
\tablerule\tablepad
&& \#  {\rm Independent\  Variables} 
&&\half N(N+1) &&Nd + d - \half d(d+1)&&\cr
\tablerule
\noalign{\bigskip}
\noalign{\narrower\noindent{ } }
 }}$$

For $N\leq d$ a solution to \conds\ will exist \bracket.
Evidentally, for $N>d$, a given current $J$ can only give relations
between amplitudes for values of kinematic invariants that
lie on a codimension $N-d$ subvariety of the manifold of
allowed invariants. Of course, an infinite number of subvarieties
will be covered by using the infinite number of currents $J$
at different mass levels.
In conjunction with analyticity and growth conditions
these relations might well fully determine the amplitudes.
Unfortunately,  this is difficult to prove, except in
special cases (e.g. $4$-particle scattering in $d=3$).

H. Verlinde and E. Witten pointed out  an error in our
first attempt to circumvent this problem.

\newsec{Fix}

We simply embed the CFT $\CC_1$ of $\IR^{1,25}$
into a larger CFT $\CC_2$ in such a way that the
cohomology and amplitudes of $\CC_1$ are embedded in
those of $\CC_2$.  This can be done, e.g.,  by taking
$\CC_2 = \CC(\IR^{1+E,25+E})\otimes_{i=1}^E
\bigl[\langle \xi_i,\eta_i\rangle \cap \ker (\oint \eta_i)\bigr]$,
where $ \langle \xi_i,\eta_i\rangle$ are anticommuting $(0,1)$-systems,
and
$ Q_2=\oint c (T^{26+2E} +T^{\xi,\eta}) + c\p c b$
is  the differential.
\foot{The choice of signature of the additional
space is unimportant to our present considerations
because we work with complex momenta. It
is preferred from other points of view.
Indeed, H. Ooguri and C. Vafa noted that , with the above
choice of signature, if  $v_i\in \IR^{1+E,25+E}$ satisfy
$v_i\cdot v_j=0$ then we can add $Q_2\to Q_2+\Delta Q$,
with $\Delta Q= \sum_{j=1}^E i \oint   \xi_j v_j\cdot \p X $
such that $\CH_Q(\CC_1) \cong \CH_{Q_2+\Delta Q}(\CC_2)$.
}
If $\iota: \CC_1\to \CC_2$ is the embedding then
$\iota Q_1 =Q_2 \iota$, so we have a map on
cohomology, $\iota_*:\CH_Q(\CC_1) \hookrightarrow \CH_{Q_2}(\CC_2)$.

Scattering amplitudes in the theory $\CC_2$ are defined by
integrated vertex-operator correlators, with an insertion of
$\prod_{i=1}^E \xi(z_i)$ to soak up the $\xi_i, \eta_i $ zero-modes.
As in the standard Friedan-Martinec-Shenker construction
of covariant superstring amplitudes, the density is independent
of the location of the points
$z_i$ for amplitudes of states in the ``little Hilbert space''
defined by $\oint \eta_i=0$. For the same reason, the amplitude
is a function on the cohomology $\CH_{Q_2}(\CC_2)$.
It follows from the integral representation for the correlators
that  if $V_i\in \CH_{Q_1}(\CC_1)$ then
\eqn\ampseq{
\CA(V_1, \dots, V_N)_{\CC_1} =
\CA(\iota_*(V_1),  \dots, \iota_*(V_N))_{\CC_2}
\qquad .
}
A corollary of \ampseq\  is that  $Ker(\iota_*)=0$.
\ampseq\  will not hold at higher genus.

Consider the BRST classes in  $\CH_2$:
\eqn\tachphot{
\eqalign{
c e^{i p\cdot X} & \qquad p^2=2\cr
i c  \zeta\cdot \p X e^{i q\cdot X} & \qquad q^2 = \zeta\cdot q =0\cr}
}
where $p,q,\zeta\in\IC^{26+2E}$.
Let $\CH'\subset \CH_2$ be the smallest
subspace containing these classes and invariant under the bracket.
We have
$\CH_Q(\CC_1) \hookrightarrow \CH'\hookrightarrow \CH_{Q_2}(\CC_2)$.
The first inclusion follows from  the no-ghost theorem \bracket. Both
inclusions
are proper.
\bigskip
\noindent
Lemma.  $\CA': (\CH')^{\otimes n} \to \IC$
satisfies axioms P1,P2,P3.

\noindent
{\it Proof}:
P1 is a consequence of the $\CP_{26+2E}$ invariance of
$\CH'$ and of the OPE.
P2: Although  $\CH'$ is a proper subspace of
 $\CH_2$,  since $\CH'$ is closed under bracket we will not generate any
states outside this space by factorization.
P3  may be proved from the integral representation for the
amplitudes in exactly the same way as for the amplitudes of $\CC_1$.
$\spadesuit$

In $\CH'$ there is ``enough room'' to solve \conds\ for $N\leq 26+2E$.
Therefore, we can apply the same argument as in
\bracket\ to conclude that $\CA'$ is uniquely fixed up to
a constant $c_N$.
Now,  $\CH\hookrightarrow \CH'$, so by \ampseq\
 the ``physical'' amplitudes $\CA: (\CH)^{\otimes N}\to\IC$
are determined  {\it a fortiori}.
Of course, $E$ is
arbitrary so the argument applies to all $N$.  Thus, by enlarging
the set of amplitudes one finds a closed set of relations
 which determine all amplitudes.
Exactly this style of argument was used to establish the
uniqueness theorem for closed strings in \bracket.

\newsec{Conclusion}

One unsatisfactory aspect of this argument is that
we cannot adopt a purely axiomatic approach to
determining the $N>26$ amplitudes  in strict analogy
to the $N\leq 26$ case  in \bracket.  The reason is
that it is not obvious that  the
axioms on $\CA$  imply the extension $\CA'$ satisfies
the same axioms. Thus, in our Lemma we have
 to appeal to the explicit integral representations for $\CA'$
to establish axioms $P2,P3$.

In our search for a universal spacetime gauge principle
we are  led to the general construction of tensoring an
on-shell background by a $c=0$ CFT. It is intriguing that in searching
for a universal worldsheet gauge principle N. Berkovits and
C. Vafa  used a similar (albeit more intricate) construction
\ref\berkvaf{N. Berkovits and C. Vafa, ``On the Uniqueness of String
Theory,''  HUTP-93/A031, KCL-TH-93-13; hep-th/9310170}.

\bigskip
\centerline{\bf Acknowledgements}

I would like to thank T. Banks, J. Distler, B.Lian,
S. Shenker, and G. Zuckerman for useful discussions.
I would also like to thank the Rutgers Physics
Dept. for hospitality.
This work is supported by DOE grants DE-AC02-76ER03075,
DE-FG02-92ER25121, DE-FG05-900ER40559,
and by a Presidential Young Investigator Award.

\listrefs

\bye